\RequirePackage{ifpdf}
\ifpdf 
\documentclass[pdftex]{sigma}
\else
\documentclass{sigma}
\fi

\numberwithin{equation}{section}

\def\coun{\varepsilon}

\def\<{\langle}
\def\>{\rangle}
\newcommand{\CC}{{\mathbb C}}

\newcommand{\ZZ}{{\mathbb Z}}

\def\r#1{(\ref{#1})}

\def\ot{\otimes}

\newcommand{\Uqdva}{{U_{q}^{}(\widehat{\mathfrak{sl}}_{2})}}
\def\sk#1{\left(#1\right)}

\def\Pep{{P}^+_e}
\def\Pem{{P}^-_e}
\def\Pepm{{P}^\pm_e}
\def\Pemp{{P}^\mp_e}
\def\Pfpm{{P}^\pm_{f}}
\def\Pfp{{P}^+_f}
\def\Pfm{{P}^-_f}

\def\Uqgln{U_q(\widehat{\mathfrak{gl}}_N)}

\def\Uqsl2{U_q(\widehat{\mathfrak{sl}}_2)}

\def\Uqtri{U_q(\widehat{\mathfrak{gl}}_3)}

\def\EE{{\rm E}}\def\FF{{\rm F}}

\def\prodl{\mathop{\overleftarrow\prod}\limits}

\def\ct{\mathbb{T}}
\def\End{\textrm{End}}
\def\Uqbp{U_q(\mathfrak{b}^+)}
\def\bbb{\mathbb{B}}

\def\ccR{\mathbb{R}}
\def\si{\sigma}
\def\RR{{\rm R}}
\def\LL{{\rm L}}

\def\E{{\rm E}}

\def\tSym{\overline{\rm Sym}}

\def\Ff{f}
\def\Ee{e}

\def\qSym{\overline{\rm Sym}}

\def\qed{\hfill$\square$\medskip}
\def\ccc{\mathbb{C}}
\def\Seb{{\sf E}_2}
\def\Sfb{{\sf F}_2}
\def\stackreb#1#2{\mathrel{\mathop{#2}\limits_{#1}}}
\def\res#1{\stackreb{#1}{\mbox{\rm res}}}
\def\simm{\thickapprox}
\def\simJ{\sim_{\!\scriptscriptstyle{J}}}
\def\simI{\sim_{\!\scriptscriptstyle{I}}}
\def\simK{\sim_{\!\scriptscriptstyle{K}}}

\def\Scp{\mathcal{S}}
\def\CalE{\mathcal{E}}
\def\CalF{\mathcal{F}}
\def\KerE{\mathbb{E}}
\def\KerF{\mathbb{F}}

\begin{document}

\allowdisplaybreaks

\renewcommand{\thefootnote}{$\star$}

\renewcommand{\PaperNumber}{094}

\FirstPageHeading

\ShortArticleName{Universal Bethe Ansatz and Scalar Products of Bethe Vectors}

\ArticleName{Universal Bethe Ansatz and
Scalar Products\\ of Bethe Vectors\footnote{This paper is a
contribution to the Proceedings of the International Workshop ``Recent Advances in Quantum Integrable Systems''. The
full collection is available at
\href{http://www.emis.de/journals/SIGMA/RAQIS2010.html}{http://www.emis.de/journals/SIGMA/RAQIS2010.html}}}

\Author{Samuel BELLIARD~$^\dag$, Stanislav PAKULIAK~$^\ddag$ and Eric RAGOUCY~$^\S$}

\AuthorNameForHeading{S. Belliard, S. Pakuliak and E. Ragoucy}

\Address{$^\dag$~Istituto Nazionale di Fisica Nucleare, Sezione di Bologna, Italy}
\EmailD{\href{mailto:belliard@bo.infn.it}{belliard@bo.infn.it}}

\Address{$^\ddag$~Institute of Theoretical \& Experimental Physics, 117259 Moscow, Russia\\[0.5ex]
\hphantom{$^\ddag$}~Laboratory of Theoretical Physics, JINR, 141980 Dubna, Moscow reg., Russia\\[0.5ex]
\hphantom{$^\ddag$}~Moscow Institute of Physics and Technology, 141700, Dolgoprudny, Moscow reg., Russia}
\EmailD{\href{mailto:pakuliak@theor.jinr.ru}{pakuliak@theor.jinr.ru}}

\Address{$^\S$~Laboratoire de Physique Th\'eorique LAPTH,
CNRS and Universit\'e de Savoie, \\
\hphantom{$^\S$}~BP 110, 74941 Annecy-le-Vieux Cedex, France}
\EmailD{\href{mailto:eric.ragoucy@lapp.in2p3.fr}{eric.ragoucy@lapp.in2p3.fr}}

\ArticleDates{Received October 25, 2010;  Published online December 14, 2010}

\Abstract{An integral presentation for the scalar products of  nested Bethe vectors
for the quantum integrable
models associated with the quantum af\/f\/ine algebra $U_q(\widehat{\mathfrak{gl}}_3)$ is given. This result is obtained in the
framework of the universal Bethe ansatz, using presentation of the universal Bethe vectors in terms
of the total currents of a ``new'' realization of the quantum af\/f\/ine algebra $U_q(\widehat{\mathfrak{gl}}_3)$.}


\Classification{17B37; 81R50}

\renewcommand{\thefootnote}{\arabic{footnote}}
\setcounter{footnote}{0}

\section{Introduction}

\looseness=1
The problem of computing correlation functions is one of the most challenging problem
in the f\/ield of quantum integrable models, starting from the establishment of the Bethe ansatz method
in \cite{Be}. For the models where algebraic Bethe ansatz \cite{TF,FS,FST,F} is
applicable, this problem can be reduced to the calculation of  scalar products of of\/f-shell
Bethe vectors. These latters are Bethe vectors where the Bethe parameters are not constrained
to obey the Bethe Ansatz equations anymore.
For $\mathfrak{gl}_2$-based integrable models,
these scalar products were calculated in \cite{Ko,KoIz,I-dwbc} and are given by the sums over partitions of
two sets of the Bethe parameters. Lately, it was shown by N. Slavnov \cite{Sla}, that if one set
of Bethe parameters satisf\/ies Bethe equations (which
guarantees that the Bethe vectors
are eigenvectors of the transfer matrix), then the formula for scalar products can be
written in a determinant form. This form is very useful to get an integral presentation for
correlation functions \cite{KKS,IKMT,KMT,Mai-gr} in the thermodynamic limit.

There is a wide class of quantum integrable models associated with the algebra $\mathfrak{gl}_N$ ($N>2$).
An algebraic Bethe ansatz for these type models is called hierarchical
(or nested) and was introduced by
P. Kulish and N. Reshetikhin \cite{KR}. This method is based on a
recursive procedure which reduces the eigenvalue
problem for the transfer matrix for the model with $\mathfrak{gl}_N$ symmetry
to an analogous problem
for the model with $\mathfrak{gl}_{N-1}$ symmetry. Assuming that the problem
for $N=2$ is solved, this method
allows to f\/ind hierarchical Bethe equations. Explicit formulas for the
hierarchical Bethe vectors in terms of the matrix elements of the $
\mathfrak{gl}_N$ monodromy
matrix can be found in \cite{BR}, but these complicated expressions are
very dif\/f\/icult to handle.

The solution of this problem, namely the formulas for the of\/f-shell Bethe vectors in terms of monodromy
matrix, was found in \cite{VT}. These vectors are called universal, because they have the same
structure for the dif\/ferent models sharing the same hidden symmetry.
 This construction requires a very complicated procedure of calculation
 of
the trace of  projected tensor powers of the monodromy matrix. It was performed
in \cite{VTcom}, but
only on the level of the evaluation representation of $\Uqgln$ monodromy matrix.

There is a new, alternative approach to the construction of universal Bethe vectors for~$\mathfrak{gl}_N$
symmetry models
 using current realizations of the
quantum af\/f\/ine algebras \cite{D88} and \mbox{using} a~Ding--Frenkel isomorphism between current and
$\LL$-operators realizations
of the quantum af\/f\/ine algebra $\Uqgln$
\cite{DF}. This approach allows to obtain explicit formulas for
the universal Bethe vectors in terms of the current generators of the quantum af\/f\/ine algebra
$\Uqgln$ for
arbitrary highest weight representations.
It was proved in \cite{KP-GLN}  that the
two methods of construction of the universal Bethe vectors coincide
on the level of the evaluation representations.
Furthermore,  it was shown in \cite{FKPR} that the eigenvalue property of the hierarchical universal
Bethe vectors can be reformulated as a problem of ordering of the current generators in the product
of the universal transfer matrix and the universal Bethe vectors. It was
proved that the eigenvalue property
appears only if parameters of the universal Bethe vectors satisfy the universal Bethe equations of the
analytical Bethe ansatz \cite{ACDFR}.

The universal Bethe vectors in terms of current generators have an integral presentation,
as an integral
transform with some kernel of the product of the currents.
In the $\Uqdva$ case, this integral representation produces immediately
an integral formula for the
scalar product of of\/f-shell Bethe vectors \cite{Sl-rev} which is equivalent to the
Izergin--Korepin formula.
In this article, we present an integral presentation of the universal of\/f-shell Bethe vectors based on
the quantum af\/f\/ine algebra $\Uqtri$. These integral formulas lead to  integral formulas for
scalar products with some kernel. The corresponding formula \r{sc1} is
the main result of our paper.
The problem left to be done is to transform the integral form we have obtained to a determinant form
which can be very useful for the application
to quantum integrable models associated to $\mathfrak{gl}_N$ symmetry algebra.

\section{Universal Bethe vectors in terms of L-operator}\label{BV-Lop}

\subsection[$U_q(\widehat{\mathfrak{gl}}_3)$ in L-operator formalism]{$\boldsymbol{U_q(\widehat{\mathfrak{gl}}_3)}$ in L-operator formalism}

Let $\E_{ij}\in{\textrm{End}}(\CC^3)$ be a matrix with the only nonzero entry
equals to $1$
at the intersection of the $i$-th row and $j$-th column.
Let $\RR(u,v)\in{\textrm{End}}(\CC^3\ot\CC^3)\ot \CC[[{v}/{u}]]$,
\begin{gather*}
\RR(u,v)  =  \sum_{1\leq i\leq 3}\E_{ii}\ot \E_{ii}  +  \frac{u-v}{qu-q^{-1}v}
\sum_{1\leq i<j\leq 3}(\E_{ii}\ot \E_{jj}+\E_{jj}\ot \E_{ii})\nonumber
\\
\phantom{\RR(u,v)  = }{} +   \frac{q-q^{-1}}{qu-q^{-1}v}\sum_{1\leq i<j\leq 3}
(u \E_{ij}\ot \E_{ji}+ v \E_{ji}\ot \E_{ij})
\end{gather*}
be a trigonometric $\RR$-matrix associated with the vector
representation of ${\mathfrak{gl}}_3$. Let $q$ be  a complex parameter
dif\/ferent from zero or a root of unity.

The algebra $\Uqtri$ (with zero central charge and the gradation operator
dropped out) is an associative algebra with unit, generated by the modes
$\LL^{\pm}_{i,j}[\pm k]$, $k\geq 0$, $1\leq i,j\leq 3$ of the
$\LL$-operators
$\LL^{\pm}(z)=\sum\limits_{k=0}^\infty\sum\limits_{i,j=1}^3 \E_{ij}\otimes \LL^{\pm}_{i,j}[\pm k]z^{\mp
k}$, subject to the relations
\begin{gather}\label{L-op-com}
\begin{gathered}
 \RR(u,v)\cdot (\LL^{\pm}(u)\ot \mathbf{1})\cdot (\mathbf{1}\ot \LL^{\pm}(v))=
(\mathbf{1}\ot \LL^{\pm}(v))\cdot (\LL^{\pm}(u)\ot \mathbf{1})\cdot \RR(u,v)
 ,\\
\RR(u,v)\cdot (\LL^{+}(u)\ot \mathbf{1})\cdot (\mathbf{1}\ot \LL^{-}(v))=
(\mathbf{1}\ot \LL^{-}(v))\cdot (\LL^{+}(u)\ot \mathbf{1})\cdot \RR(u,v)
,
\end{gathered}
\\
\label{L-op-com1}
\LL^{+}_{j,i}[0]=\LL^{-}_{i,j}[0]=0,
\qquad 1\leq i<j \leq 3 ,
\\
\label{L-op-com2}
\LL^{+}_{k,k}[0]\LL^{-}_{k,k}[0]=1,
\qquad 1\leq k\leq 3  .
\end{gather}
Actually, we will not impose the condition \r{L-op-com2} since the universal
Bethe vectors will be constructed only from one $\LL$-operator, say
$\LL^+(z)$.

Subalgebras formed by the modes $\LL^\pm[n]$ of the $\LL$-operators
$\LL^\pm(z)$
are the standard  Borel subalgebras $U_q(\mathfrak{b}^\pm)\subset\Uqgln$.
These Borel subalgebras are Hopf subalgebras of $\Uqgln$.
Their coalgebraic structure is given by the formulae
\begin{gather*}
\Delta \big(\LL^{\pm}_{i,j}(u)\big)=\sum_{k=1}^N\ \LL^{\pm}_{k,j}(u)\otimes
\LL^{\pm}_{i,k}(u) .
\end{gather*}

\subsection{Universal of\/f-shell Bethe vectors}

We will follow the construction of the of\/f-shell Bethe vectors due to
 \cite{VT}. Let  $\LL^{}(z)=\sum\limits_{k=0}^\infty \sum\limits_{i,j=1}^3 \E_{ij}\otimes
\LL^{}_{i,j}[ k]z^{-k}$ be the $\LL$-operator\footnote{We omit
superscript + in this $\LL$-operator, since will consider only
positive standard Borel subalgebra ${U}_q({\mathfrak{b}}^+)$ here and
below.} of the Borel
subalgebra ${U}_q({\mathfrak{b}}^+)$ of $\Uqtri$ satisfying the Yang--Baxter commutation relation with a
$R$-matrix $\RR(u,v)$.  We use the notation $\LL^{(k)}(z)\in \sk{\CC^3}^{\otimes
M}\ot {U}_q({\mathfrak{b}}^+)$ for
$\LL$-operator acting nontrivially on $k$-th
tensor factor in the product $\sk{\CC^{3}}^{\otimes M}$ for $1\leq k\leq M$.
 Consider a series in $M$ variables
\begin{gather}\label{product}
\ct(u_1,\ldots,u_M)=\LL^{(1)}(u_1)\cdots \LL^{(M)}(u_M)
\cdot \ccR^{(M,\ldots,1)}(u_M,\ldots,u_1) ,
\end{gather}
with coef\/f\/icients in $\sk{\End(\CC^3)}^{\ot M}\ot {U}_q({\mathfrak{b}}^+)$,
where
\begin{gather}\label{Rproduct}
\ccR^{(M,\ldots,1)}(u_M,\ldots,u_1)=
{\prodl_{M \geq j > 1}} \ {\prodl_{j > i \geq 1}} {\RR}^{(ji)}(u_j,u_i) .
\end{gather}
In the ordered product of $R$-matrices \r{Rproduct}, the ${\RR}^{(ji)}$
factor is on
the left of the ${\RR}^{(ml)}$ factor if $j>m$, or $j=m$ and $i>l$.
Consider the set of  variables
\begin{gather*}
\{\bar{t},\bar s\} = \left\{
t_{1},\ldots,t_{a}; s_{1},\ldots,
s_{b}\right\} .
\end{gather*}
Following \cite{VT}, let
\begin{gather}
\bbb(\bar t,\bar s)
=  \prod_{1\leq j\leq b}\    \prod_{1\leq i\leq a}
\frac{qs_j-q^{-1}t_i}{s_j-t_i}\nonumber\\
\phantom{\bbb(\bar t,\bar s)=}{} \times({\rm tr}_{(\CC^3)^{\ot(a+b)}}\ot {\rm id})\ \big(\ct(t_1,\ldots,t_{a};
s_1,\ldots,s_{b})\E_{21}^{\ot a}\ot \E_{32}^{\ot b}\ot 1\big) .\label{Vit-el}
\end{gather}
The element $\ct(\bar t,\bar s)$
in \r{Vit-el} is given by \r{product}
with obvious identif\/ication. The coef\/f\/icients of
$\bbb(\bar t,\bar s)$ are elements of the Borel subalgebra
$U_q({\mathfrak{b}}^+)$.

We call  vector $v$ \emph{a right weight singular vector} if it is annihilated
by any positive mode  $\LL_{i,j}[n]$, $i>j$, $n\geq 0$ of the matrix elements
of the $\LL^+(z)$ operator and is
an eigenvector of the diagonal matrix entries
$\LL^+_{i,i}(z)$:
\begin{gather}\label{hwv}
\LL^{+}_{i,j}(z)  v=0 ,\quad i>j ,\qquad \LL^{+}_{i,i}(z)  v=\lambda_i(z)  v ,\quad
i=1,\ldots,3 .
\end{gather}
For any right $\Uqtri$-module $V$ with a
right singular vector $v$, denote{\samepage
\begin{gather}\label{Vel-vs-pr}
\bbb_V(\bar t,\bar s)=\bbb(\bar t,\bar s)v.
\end{gather}
The vector valued function ${\mathbb{B}}_{V}(\bar t,\bar s)$
was called in  \cite{VT,VTcom} {\it universal off-shell Bethe vector}.}

We call  vector $v'$ \emph{a left weight singular vector} if it is annihilated
by any positive mode  $\LL_{i,j}[n]$, $i<j$, $n\geq 0$ of the matrix elements
of the $\LL$-operator $\LL^+(z)$  and is
an eigenvector of the diagonal matrix entries
$\LL^+_{i,i}(z)$:
\begin{gather}\label{hwvl}
0=v' \LL^{+}_{i,j}(z) ,\quad i<j ,\qquad v'  \LL^{+}_{i,i}(z)=\mu_i(z)  v' ,\quad
i=1,\ldots,3 .
\end{gather}
For any left $\Uqtri$-module $V'$ with a
left singular vector $v'$, denote
\begin{gather}\label{Vel-vs-prl}
\ccc_{V'}(\bar \tau,\bar \sigma)=v'\ccc(\bar \tau,\bar \sigma)  ,
\end{gather}
where
\begin{gather*}
\{\bar{\tau},\bar \sigma\} = \left\{
\tau_{1},\ldots,\tau_{a}; \sigma_{1},\ldots,
\sigma_{b}\right\} ,
\end{gather*}
and
\begin{gather*}
\ccc(\bar \tau,\bar \sigma)
=  \prod_{1\leq j\leq b}\    \prod_{1\leq i\leq a}
\frac{q\sigma_j-q^{-1}\tau_i}{\sigma_j-\tau_i}\nonumber\\
\phantom{\ccc(\bar \tau,\bar \sigma)=}{} \times({\rm tr}_{(\CC^3)^{\ot(a+b)}}\ot {\rm id})\ \big(\ct(\tau_1,\ldots,\tau_{a};
\sigma_1,\ldots,\sigma_{b})\E_{12}^{\ot a}\ot \E_{23}^{\ot b}\ot
1\big) .
\end{gather*}

Our goal is to calculate the scalar product
\begin{gather}\label{scal-prod}
\langle \ccc_{V'}(\bar \tau,\bar \sigma),\bbb_V(\bar t,\bar s)\rangle .
\end{gather}
There is a direct way to solve this problem, using the exchange relations of the
$\LL$-operators matrix elements
 and the def\/initions of the singular weight vectors. However, this approach
 is a highly
complicated combinatorial problem. Instead, we will use another presentation
of the universal Bethe vectors
given recently in the paper  \cite{KP-GLN,OPS}, using current realization
of the quantum af\/f\/ine algebra $\Uqtri$ and method of projections introduced in~\cite{ER} and developed in~\cite{EKhP}.

\section[Current realization of  $U_q(\widehat{\mathfrak{gl}}_3)$]{Current realization of  $\boldsymbol{U_q(\widehat{\mathfrak{gl}}_3)}$}

\subsection{Gauss decompositions of L-operators}

The relation between the $\LL$-operator realization of $\Uqtri$ and its current realization \cite{D88}
is known since the work \cite{DF}.
To build an isomorphism between these two realizations, one has to consider
the Gauss decomposition of the $\LL$-operators and identif\/ies linear combinations
of some
Gauss coordinates with the total currents of $\Uqtri$ corresponding to the simple roots of
$\mathfrak{gl}_3$.
Recently, it was shown in \cite{OPS} that there are two dif\/ferent but isomorphic current realization
of $\Uqtri$. They correspond to dif\/ferent embeddings of smaller algebras into bigger ones and to dif\/ferent
type of Gauss decompositions of the fundamental $\LL$-operators.
These two dif\/ferent current realizations have dif\/ferent commutation
relations, dif\/ferent current
comultiplications and dif\/ferent associated projections onto intersections of the current and
Borel subalgebras
of~$\Uqtri$.

Our way to calculate the scalar product of Bethe vectors \r{scal-prod}
is to use an alternative form
to expressions \r{Vel-vs-pr} and \r{Vel-vs-prl} for the universal Bethe
vectors. It is written in terms of
projections of products of currents onto intersections of the current and Borel subalgebras
of $\Uqtri$. In this case, the universal Bethe vectors can be written
as some integral  and the calculation
of the scalar product is reduced to the calculation of an integral of some rational function.

For the $\LL$-operators f\/ixed by the relations \r{L-op-com} and
\r{L-op-com1}, we consider the
following decompositions into Gauss coordinates $\FF^{\pm}_{j,i}(t)$,
$\EE^{\pm}_{i,j}(t)$, $j>i$ and $k^\pm_{i}(t)$:
\begin{gather}\label{GF2}
\LL^{\pm}_{i,j}(t)  =  \FF^{\pm}_{j,i}(t)k^+_{i}(t)+\sum_{1\leq m< i}
\FF^{\pm}_{j,m}(t)k^\pm_{m}(t)\EE^{\pm}_{m,i}(t) ,\qquad 1\leq i<j\leq 3 ,\\
\label{GK2}
\LL^{\pm}_{i,i}(t)  =  k^\pm_{i}(t) +\sum_{1\leq m< i} \FF^{\pm}_{i,m}(t)k^\pm_{m}(t)
\EE^{\pm}_{m,i}(t) ,\qquad i=1,2,3,\\
\label{GE2}
\LL^{\pm}_{i,j}(t)  =  k^\pm_{j}(t)\EE^{\pm}_{j,i}(t)+\sum_{1\leq m<j}
\FF^{\pm}_{j,m}(t)k^\pm_{m}(t)\EE^{\pm}_{m,i}(t) ,\qquad   3\geq i>j\geq 1.
\end{gather}

Using the arguments of \cite{DF}, we may obtain for the linear combinations
of the Gauss coordinates $(i=1,2)$
\begin{gather*}
F_i(t)=\FF^{+}_{i+1,i}(t)-\FF^{-}_{i+1,i}(t) ,\qquad
E_i(t)=\EE^{+}_{i,i+1}(t)-\EE^{-}_{i,i+1}(t) ,
\end{gather*}
and for the Cartan currents $k^\pm_i(t)$,
 the commutation relations of the quantum af\/f\/ine algebra
$\Uqtri$ with zero
central charge and the gradation operator dropped out.  In terms of the total currents
$F_i(t)$, $E_i(t)$ and of the Cartan currents $k^\pm_i(t)$, these commutation
relations read
\begin{gather}
\label{gln-com1}
(q^{}z-q^{-1}w)E_{i}(z)E_{i}(w)  =  E_{i}(w)E_{i}(z)(q^{-1}z-q^{}w)  , \\
(q^{-1}z-qw)E_{i}(z)E_{i+1}(w)  =  E_{i+1}(w)E_{i}(z)(z-w)  , \\
k_i^\pm(z)E_i(w)\left(k_i^\pm(z)\right)^{-1}  = \frac{z-w}{q^{-1}z-q^{}w}E_i(w) , \\
k_{i+1}^\pm(z)E_i(w)\left(k_{i+1}^\pm(z)\right)^{-1}  =
 \frac{z-w}{q^{}z-q^{-1}w}E_i(w)  ,  \\
k_i^\pm(z)E_j(w)\left(k_i^\pm(z)\right)^{-1}  =  E_j(w), \qquad {\rm if}\quad i\not=j,j+1  ,
\\
(q^{-1}z-q^{}w)F_{i}(z)F_{i}(w)  =  F_{i}(w)F_{i}(z)(q^{}z-q^{-1}w)  , \\
(z-w)F_{i}(z)F_{i+1}(w)  =  F_{i+1}(w)F_{i}(z)(q^{-1}z-qw)  ,  \\
k_i^\pm(z)F_i(w)\left(k_i^\pm(z)\right)^{-1}  = \frac{q^{-1}z-qw}{z-w}F_i(w)  , \\
k_{i+1}^\pm(z)F_i(w)\left(k_{i+1}^\pm(z)\right)^{-1}  =
 \frac{q^{}z-q^{-1}w}{z-w}F_i(w)  , \\
k_i^\pm(z)F_j(w)\left(k_i^\pm(z)\right)^{-1}  =  F_j(w), \qquad {\rm if}\quad i\not=j,j+1 , \\
[E_{i}(z),F_{j}(w)]  =  \delta_{{i},{j}}  \delta(z/w)
(q-q^{-1})\big(k^-_{i+1}(z)/k^-_{i}(z)-k^+_{i+1}(w)/k^+_{i}(w)\big)  ,
\label{gln-comF}
\end{gather}
with Serre relations
\begin{gather*}
{\rm Sym}_{z_1,z_{2}}
\Big(E_{i}(z_1)E_{i}(z_2)E_{i\pm 1}(w)
 -(q+q^{-1})E_{i}(z_1)E_{i\pm 1}(w)E_{i}(z_2) \nonumber\\
\qquad{} +E_{i\pm 1}(w)E_{i}(z_1)E_{i}(z_2)\Big)=0  ,\nonumber\\
{\rm Sym}_{z_1,z_{2}}
\Big(F_{i}(z_1)F_{i}(z_2)F_{i\pm 1}(w)
 -(q+q^{-1})F_{i}(z_1)F_{i\pm 1}(w)F_{i}(z_2) \nonumber\\
\qquad{} +F_{i\pm 1}(w)F_{i}(z_1)F_{i}(z_2)\Big)=0  .
\end{gather*}
Formulae \eqref{gln-com1}--\eqref{gln-comF} should be considered as
formal series identities describing the inf\/inite set of relations between modes of the
currents. The symbol $\delta(z)$ entering these relations is a formal series $\sum\limits_{n\in\ZZ} z^n$.

\subsection{Borel subalgebras and projections on their intersections}
\label{cBs}

We consider two types of Borel subalgebras in the
algebra $\Uqtri$. Borel subalgebras $U_q(\mathfrak{b}^\pm)\subset \Uqtri $
are generated by the modes of the $\LL$-operators $\LL^{(\pm)}(z)$ respectively.
For the generators in these subalgebras, we can use instead modes of the Gauss coordinates
\r{GF2}--\r{GE2},
$\EE^\pm_{i,i+1}(t)$, $\FF^\pm_{i+1,i}(t)$, $k^\pm_{j}(t)$.

Other types of Borel subalgebras are related to the current realizations of
$\Uqgln$ given in the previous subsection.
We consider f\/irst the  current Borel subalgebras generated by the modes
of the currents $E_i(t)$, $F_i(t)$, $k^\pm_j(t)$.

 The Borel subalgebra $U_F\subset \Uqtri$ is generated by
modes of the currents
$F_i[n]$, $k^+_j[m]$, $i=1,2$, $j=1,2,3$, $n\in\ZZ$
and $m\geq0$. The Borel subalgebra
$U_E\subset \Uqtri$ is generated by modes of the currents
$E_i[n]$, $k^-_j[-m]$, $i=1,2$, $j=1,2,3$, $n\in\ZZ$ and
$m\geq0$. We will consider also the subalgebra $U'_F\subset U_F$, generated
by the elements
$F_i[n]$, $k^+_j[m]$, $i=1,2$, $j=1,2,3$, $n\in\ZZ$
and $m>0$, and the subalgebra $U'_E\subset U_E$ generated by
the elements
$E_i[n]$, $k^-_j[-m]$, $i=1,2$, $j=1,2,3$, $n\in\ZZ$
and $m>0$.
In the following, we will be interested in the intersections
\begin{gather*}
U_f^- = U'_F\cap U_q(\mathfrak{b}^-) ,\qquad
U_F^+=U_F\cap U_q(\mathfrak{b}^+) ,
\\
U_E^- = U_E\cap U_q(\mathfrak{b}^-) ,\qquad
U_e^+=U'_E\cap U_q(\mathfrak{b}^+) ,
\end{gather*}
and will describe properties of the projections on these intersections.
We call $U_F$ and $U_E$ {\em the current Borel subalgebras}.

In \cite{D88} the \emph{current} Hopf structure for the algebra
$\Uqtri$ has been def\/ined as:
\begin{gather}
\Delta^{(D)}\sk{E_i(z)} =1\ot E_i(z) + E_i(z)\ot k^-_{i+1}(z)\sk{k^-_{i}(z)}^{-1},\nonumber\\
\Delta^{(D)}\sk{F_i(z)} =F_i(z)\ot 1 +
k^+_{i+1}(z)\sk{k^+_{i}(z)}^{-1}\ot F_i(z),\label{gln-copr}\\
\Delta^{(D)}\sk{k^\pm_i(z)} =k^\pm_i(z)\ot k^\pm_{i}(z).\nonumber
\end{gather}
With respect to the current Hopf structure, the current Borel subalgebras are Hopf
subalgebras of $\Uqtri$. One may construct the whole algebra $\Uqtri$ from one of its current
Borel subalgebras using the current Hopf structure and the Hopf pairing
\begin{gather}
\langle E_i(z),F_j(w)\rangle  = (q-q^{-1})\delta_{ij}\delta(z/w),\nonumber\\
\langle \psi_{i}^-(z),k_{i+1}^+(w)\rangle =\langle k_{i}^-(z),\psi_{i}^+(w)\rangle^{-1}=
\frac{q-q^{-1}z/w}{1-z/w} ,\label{Hopf-pair}\\
\langle \psi_{i}^-(z),k_{i}^+(w)\rangle =\langle k_{i+1}^-(z),\psi_{i}^+(w)\rangle^{-1}=
\frac{q^{-1}-qz/w}{1-z/w} ,\nonumber
\end{gather}
where
\begin{gather}\label{Car-com}
\psi_i^\pm(t)=k^\pm_{i+1}(t)\sk{k_i^\pm(t)}^{-1},\qquad i=1,2 .
\end{gather}
Formulas \r{Hopf-pair} can be obtained from the commutation relations
\r{gln-com1}--\r{gln-comF}
using the commutator rules \r{com-qd} in the quantum double.

One can check \cite{EKhP,KPT} that  the intersections  $U_f^-$
and $U_F^+$, respectively  $U_e^+$
and $U_E^-$, are subalgebras and coideals with respect to the Drinfeld
coproduct \r{gln-copr}:
\begin{gather*}
\Delta^{(D)}(U_F^+)\subset  U_F^+\ot\Uqtri,\qquad
\Delta^{(D)}(U_f^-)\subset \Uqtri\ot U_f^- ,
\\
\Delta^{(D)}(U_e^+)\subset  U_e^+\ot\Uqtri,\qquad
\Delta^{(D)}(U_E^-)\subset \Uqtri\ot U_E^- ,
\end{gather*}
and that the multiplication $m$ in $\Uqtri$ induces an isomorphism of vector spaces
\[
m: \ U_f^-\ot U_F^+\to U_F ,\qquad m: \ U_e^+\ot U_E^-\to U_E.
\]
According to the general theory presented in \cite{EKhP}, we
introduce
the projection operators
\begin{gather*}
\Pfp:\ U_F\subset \Uqtri \to U_F^+,\qquad
\Pfm:\ U_F\subset \Uqtri \to U_f^-, \nonumber\\
\Pep:\ U_E\subset \Uqtri \to U_e^+ ,\qquad
\Pem: \ U_E\subset \Uqtri \to U_E^- .
\end{gather*}
They are respectively def\/ined by the prescriptions
\begin{gather}\label{pgln}
\Pfp(\Ff_-\ \Ff_+) = \coun(\Ff_-)\ \Ff_+, \qquad
\Pfm(\Ff_-\ \Ff_+)=\Ff_-\ \coun(\Ff_+),
\quad \forall\, \Ff_-\in U_f^-,
\quad \forall\,\Ff_+\in U_F^+  ,
\\
\label{pgln2}
\Pep(\Ee_+\ \Ee_-) = \Ee_+\ \coun(\Ee_-), \qquad
\Pem(\Ee_-\ \Ee_+)=\coun(\Ee_+)\ \Ee_-,
\quad \forall\,\Ee_-\in U_E^-,
\quad \forall\,\Ee_+\in U_e^+  ,
\end{gather}
where $\coun:\Uqtri\to\CC$ is a counit map.

It was proved in \cite{KP} that the projections $\Pfpm$ and $\Pepm$ are adjoint with respect to the
Hopf pairing \r{Hopf-pair}
\[
\langle e ,\Pfpm (f) \rangle=  \langle \Pemp(e) ,f \rangle.
\]

Denote by  $\overline U_F$ an extension of the algebra $U_F$ formed
by inf\/inite sums of monomials that are ordered products
$a_{i_1}[n_1]\cdots a_{i_k}[n_k]$ with $n_1\leq\cdots\leq n_k$,
where  $a_{i_l}[n_l]$ is either $F_{i_l}[n_l]$ or $k^+_{i_l}[n_l]$.
Denote by  $\overline U_E$ an extension of the algebra $U_E$ formed
by inf\/inite sums of monomials that are ordered products
$a_{i_1}[n_1]\cdots a_{i_k}[n_k]$ with $n_1\geq\cdots\geq n_k$,
where  $a_{i_l}[n_l]$ is either $E_{i_l}[n_l]$ or $k^-_{i_l}[n_l]$.
It was proved in \cite{EKhP} that
\begin{itemize}\itemsep=0pt
\item[(1)] the action of the projections \r{pgln} can be extended to the
 algebra
$\overline U_F$;
\item[(2)] for any $\Ff\in \overline U_F$ with
$\Delta^{(D)}(\Ff)=\sum_i \Ff'_i\otimes \Ff''_i$ we have
$\Ff=\sum_i\Pfm(\Ff'_i)\cdot \Pfp(\Ff''_i)$;
\item[(3)] the action of the projections \r{pgln2} can be extended to the
 algebra $\overline U_E$;
\item[(4)] for any $\Ee\in \overline U_E$ with $\Delta^{(D)}(\Ee)=\sum_i \Ee'_i\otimes \Ee''_i$ we have
$\Ee=\sum_i\Pep(\Ee''_i)\cdot \Pem(\Ee'_i)$.
\end{itemize}

\subsection{Def\/inition of the composed currents}

We introduce the composed currents\footnote{Dif\/ferent def\/initions exist for composed currents,
we choose the one, such that the corresponding projections of them
coincide with the Gauss coordinates~\r{GF2}.}
$E_{1,3}(w)$ and $F_{3,1}(y)$ which are def\/ined by the formulas
\begin{gather}
E_{1,3}(w) =\oint_{C_0}\frac{dz}{z}  E_2(z) E_1(w)- \oint_{C_\infty}\frac{dz}{z}
E_1(w) E_2(z)\frac{q-q^{-1}w/z}{1-w/z} ,\nonumber\\
F_{3,1}(y) =\oint_{C_0}\frac{dv}{v}  F_1(y) F_2(v)- \oint_{C_\infty}\frac{dv}{v}
F_2(v) F_1(y)\frac{q-q^{-1}y/v}{1-y/v} ,\label{com-cur}
\end{gather}
where the contour integrals $\oint_{C_{0,\infty}}\frac{dz}{z}  g(z)$ are considered as integrals around
zero and inf\/inity points respectively.
The composed currents $E_{1,3}(w)$ and $F_{3,1}(y)$ belong to the completed
alge\-bras~$\overline U_E$ and~$\overline U_F$, respectively. Let us remind
that, according to these completions,
we have to understand the product of currents $E_2(z) E_1(w)$ as an analytical `function' without
singularities in the domain $|z|\ll|w|$. Analogously, the product $F_1(y) F_2(v)$ is an analytical
`function' in the domain $|y|\gg|v|$.
For practical calculation, the contour integrals in def\/initions \r{com-cur} can be understood as the
formal integrals of a Laurent series $g(z)=\sum\limits_{k\in\ZZ}g[k] z^{-k}$
picking up its zero mode coef\/f\/icient $g[0]$.

Deforming contours in the def\/ining formulas for the composed currents  we may rewrite them  dif\/ferently
\begin{gather}\label{residues}
E_{1,3}(w)=-\res{z=w}E_2(z)E_1(w) \frac{dz}{z} ,\qquad
F_{3,1}(y)=-\res{v=y}F_1(y)F_2(v) \frac{dv}{v} ,
\end{gather}
or
\begin{gather}
E_2(z)E_1(w) =
  \frac{q-q^{-1}w/z}{1-w/z}  E_{1}(w)E_{2}(z)+
\delta(w/z)E_{1,3}(w) ,\nonumber\\
F_{1}(y)F_{2}(v) =
  \frac{q-q^{-1}y/v}{1-y/v}  F_{2}(v)F_{1}(y)+
\delta(y/v)F_{3,1}(y) .\label{delta-terms}
\end{gather}

Formulas \r{residues} are convenient for the presentation of the composed currents as
products of simple root currents
\begin{gather}\label{com-as-prod}
E_{1,3}(w)=(q^{-1}-q)E_1(w)E_2(w) ,\qquad
F_{3,1}(y)=(q^{-1}-q)F_2(y)F_1(y) .
\end{gather}
Formulas \r{delta-terms} are convenient to calculate the commutation relation between total and
half-currents.
This will be done lately.

First, we calculate the formal integrals in the formulas \r{com-cur} to obtain
\begin{gather*}
 E_{1,3}(w) =E_2[0]E_1(w)-q^{-1} E_1(w)E_2[0]-\big(q-q^{-1}\big)\sum_{k=0}^\infty
E_1(w)E_2[-k]w^k ,\nonumber\\
 F_{3,1}(y) =F_1(y)F_2[0]-q F_2[0]F_1(y)-\big(q-q^{-1}\big)\sum_{k=1}^\infty
F_2[-k]F_1(y)y^k .
\end{gather*}
Here we used the series expansion
\begin{gather*}
\frac{q-q^{-1}w/z}{1-w/z}=q+\big(q-q^{-1}\big)\sum_{k=1}^{\infty}(w/z)^k=
q^{-1}+\big(q-q^{-1}\big)\sum_{k=0}^{\infty}(w/z)^k .
\end{gather*}
Introducing now the half-currents
\begin{gather*}
E^\pm_2(w)=\pm\sum_{k>0 \atop k\leq 0} E_2[k]w^{-k} ,\qquad
F^\pm_2(w)=\pm\sum_{k\geq0\atop k<0} F_2[k]w^{-k} .
\end{gather*}
and using the decomposition of the algebra $\Uqtri$ into its standard positive and negative
Borel subalgebras and  the def\/inition of the screening operators
\begin{gather*}
\Seb(E_1(w)) =E_2[0]E_1(w)-q^{-1} E_1(w)E_2[0] ,\nonumber\\
\Sfb(F_1(y)) =F_1(y)F_2[0]-q F_2[0]F_1(y),
\end{gather*}
we may write
\begin{gather*}
\Seb(E_1(w)) =E_{1,3}(w)+(q^{-1}-q)E_1(w)E^-_{2}(w) ,\nonumber\\
\Sfb(F_1(y)) =F_{3,1}(y)+(q^{-1}-q)F^-_{2}(y)F_1(y) ,
\end{gather*}
or
\begin{gather}\label{cc-v-so2}
\Seb(E_1(w))=(q^{-1}-q)E_1(w)E^+_{2}(w) ,\qquad
\Sfb(F_1(y))=(q^{-1}-q)F^+_{2}(y)F_1(y) .
\end{gather}
To obtain the latter relation, we used formulas \r{com-as-prod}
and relation between total and half-currents
$E_2(w)=E^+_2(w)-E^-_2(w)$ and $F_2(w)=F^+_2(w)-F^-_2(w)$.

\section{Universal Bethe vectors and projections}

The goal of this section is to obtain the representations for the left and right
universal Bethe vectors in terms of the integral transform of the products of the
total currents. This will generalize the results obtained for $U_q(\widehat{\mathfrak{sl}}_3)$
in the paper~\cite{KP}. The calculation of the scalar product after that will
be reduced to the calculation of the exchange relations between  products of
 total currents.

\subsection{Universal Bethe vectors through currents}

It was shown in the papers \cite{KP-GLN,OPS} that the universal right Bethe vectors
\begin{gather*}
\bbb_V(\bar t,\bar s)=\Pfp\sk{F_1(t_1)\cdots F_1(t_a)F_2(s_1)\cdots F_2(s_b)}\prod_{i=1}^ak_1(t_i)
\prod_{i=1}^bk_2(s_i)  v
\end{gather*}
can be identif\/ied with some projection of products of total currents.
Using the same method, we may prove that the left Bethe vectors
\begin{gather}\label{Eident}
\ccc_{V'}(\bar \tau,\bar \sigma)=v'  \prod_{i=1}^ak_1(\tau_i)
\prod_{i=1}^bk_2(\si_i) \Pep\sk{E_2(\sigma_b)\cdots E_2(\si_1)E_1(\tau_a)\cdots E_1(\tau_1)}
\end{gather}
can be also identif\/ied with projection of products of
total currents\footnote{In \r{Eident} all operators
are acting to the left onto left singular vector $v'$.}. So the problem
of calculating
the scalar product $\langle \ccc_{V'}(\bar \tau,\bar \sigma),\bbb_V(\bar t,\bar s)\rangle$
is reduced to the exchange relations between projections.
Fortunately, to perform this exchange, we have to calculate only modulo
the ideals in the algebra $\Uqtri$ which are
annihilated by the left/right singular vectors.

One could calculate these projections to present them in the form of a sum of products of
projections of simple and composed root currents (see formulas~\r{po1}
and~\r{po2} below). However,
this calculation has the same level of dif\/f\/iculty as the exchange relations of Bethe vectors in terms of
$\LL$-operators. The idea of the present paper is to rewrite projection formulas \r{po1} and \r{po2}
in terms of integrals of total simple root currents, and then to compute the exchange of products of
total currents. In this way, we will obtain an integral representation for the scalar product of the
of\/f-shell Bethe vectors. This calculation is much more easy, since the commutation relations
of the simple roots total currents are rather simple.

\subsection{Calculation of the universal of\/f-shell Bethe vectors}

Before presenting the formulas for the universal of\/f-shell Bethe vectors in terms of the
current generators, we have to introduce the following notations.
Consider the permutation group $S_n$ and its action on the formal series of $n$ variables
def\/ined, for the elementary transpositions $\sigma_{i,i+1}$, as follows
\begin{gather*}
\pi(\sigma_{i,i+1})G(t_1, \dots,t_i, t_{i+1},\dots, t_n) =
\frac{q^{-1}-q\,t_i/t_{i+1}}{q-q^{-1}\,t_i/t_{i+1}}\ G(t_1, \dots,t_{i+1}, t_i, \dots, t_n) .
\end{gather*}
The $q$-depending factor in this formula is chosen in such a way that each product
$F_a(t_1)\cdots$ $F_a(t_n)$
is invariant under this action. Summing the action over all the group of
permutations,
we obtain the operator $\tSym_{\, \bar t}=\sum\limits_{\sigma\in S_n}\pi(\sigma)$ acting as
follows
\begin{gather*}
\tSym_{\,\bar t}\,  G(\bar t) =
\sum\limits_{\si \in S_n}\prod\limits_{\substack{\ell<\ell'\\ \si(\ell)>\si(\ell')}}
\frac{q^{-1}-q\,t_{\si(\ell')}/t_{\si(\ell)}}
{q-q^{-1}\,t_{\si(\ell')}/t_{\si(\ell)}}\  G(^\sigma t).
\end{gather*}
The product is taken over all pairs $(\ell, \ell')$, such that conditions $\ell
< \ell'$ and $\si(\ell) > \si(\ell')$ are satisf\/ied simultaneously.

According to the results of the papers \cite{EKhP,KPT}, the calculation of
the universal of\/f-shell Bethe vectors
is reduced to the calculation of the projections
\begin{gather}\label{rBv}
\Pfp\sk{F_1(t_1)\cdots F_1(t_a)F_2(s_1)\cdots F_2(s_b)}
\end{gather}
for the right Bethe vectors and\footnote{For further convenience,
we will denote the spectral parameters of
the right Bethe vectors by latin symbols, and those of
the left vectors by greek ones.}
\begin{gather}\label{lBv}
\Pep\sk{E_2(\sigma_b)\cdots E_2(\si_1)E_1(\tau_a)\cdots E_1(\tau_1)}
\end{gather}
for the left Bethe vectors.
The calculation was detailed in~\cite{KP}. Here, we present the result of calculations and
 give several comments on how it was performed.

\begin{proposition}
The projections \eqref{rBv} and \eqref{lBv} are given by the series
\begin{gather}
 \Pfp\sk{F_1(t_1)\cdots F_1(t_a)F_2(s_1)\cdots F_2(s_b)}
\nonumber\\
\qquad  =   \sum_{k=0}^{\min\{a,b\}}\!\!\!\!
\frac{1}{k!(a-k)!(b-k)!}
\overline{{\rm Sym}}_{\bar t,\bar s}
\Big(
\Pfp\sk{F_1(t_{1})\cdots  F_1(t_{a-k})
F_{3,1}(t_{a-k+1})\cdots F_{3,1}(t_{a})}
 \nonumber\\
 \qquad \quad{} \times  \Pfp\sk{F_2(s_{k+1})\cdots
F_2(s_{b})}
Z(t_a,\ldots,t_{a-k+1};s_k,\ldots,s_1)
\Big)\label{po1}
\end{gather}
and\,\footnote{The ordering of the variables in the rational series in \r{po1} dif\/fers from the ordering
in the corresponding series in the paper \cite{KP}. This is because the rational series \r{rat-Y}
are def\/ined
dif\/ferently with respect to the paper~\cite{KP}.}
\begin{gather}
 \Pep\sk{E_2(\sigma_b)\cdots E_2(\si_1)E_1(\tau_a)\cdots E_1(\tau_1)}
\nonumber\\
 \quad {}=   \sum_{m=0}^{\min\{a,b\}}\!\!\!\!
\frac{1}{m!(a-m)!(b-m)!}\
\overline{{\rm Sym}}_{\bar\tau,\bar\si}\
\Big(  \Pep\sk{E_2(\si_{b})\cdots E_2(\si_{m+1})}
\label{po2}\\
 \qquad{} \times
\Pep\sk{ E_{1,3}(\tau_{a})\cdots E_{1,3}(\tau_{a-m+1}) E_1(\tau_{a-m})\cdots  E_1(\tau_{1})  }
Y(\tau_{a},\ldots,\tau_{a-m+1};\si_m,\ldots,\si_1)\Big) ,
\nonumber
\end{gather}
where
\begin{gather}
 Y(t_1,\ldots,t_n;x_1,\ldots,x_n) = \prod_{i=1}^n\frac{1}{1-x_i/t_i}
\prod_{j=1}^{i-1}\frac{q^{-1}-qx_i/t_j}{1-x_i/t_j}\nonumber\\
\phantom{Y(t_1,\ldots,t_n;x_1,\ldots,x_n)}{} =   \prod_{i=1}^n\frac{1}{1-x_i/t_i}
\prod_{j=i+1}^{n}\frac{q^{-1}-qx_j/t_i}{1-x_j/t_i} ,\label{rat-Y}\\
  Z(t_1,\ldots,t_n;x_1,\ldots,x_n) =Y(t_1,\ldots,t_n;x_1,\ldots,x_n)
 \prod_{i=1}^n\frac{x_i}{t_i} .\nonumber
\end{gather}
\end{proposition}

Note that the kernels \r{rat-Y} are def\/ined in such a way, that they have only $k$ simple poles
at the point $t_1,\ldots,t_k$ with respect to the variable $x_k$, $k=1,\ldots,n$.
These kernels appear in the integral presentation of the projections of the products of the
same simple root currents (see~\r{por4} below).

The proof of the formulas \r{po1} and \r{po2} is similar to the proof presented in the paper~\cite{KP}.
We will not repeat this calculations here, but for completeness, we collect all necessary formulas.
As a f\/irst step, we present the products of currents $F_2(s_1)\cdots F_2(s_b)$ and
$E_2(\sigma_b)\cdots E_2(\si_1)$ in a~normal ordered form using properties of the projections
given at the end of the Subsection~\ref{cBs}:
\begin{gather*}
F_2(s_1)\cdots F_2(s_b)
  =\sum_{k=0}^b  \frac{1}{k!(b-k)!} \overline{{\rm Sym}}_{\bar s}
\big(\Pfm\sk{F_2(s_1)\cdots F_2(s_k)}\cdot \Pfp\sk{F_2(s_{k+1})\cdots F_2(s_b)}\big)
,\\
E_2(\sigma_b)\cdots E_2(\si_1) \\
\qquad{}
 =\sum_{m=0}^b  \frac{1}{m!(b-m)!} \overline{{\rm Sym}}_{\bar \si}
\big(\Pep\sk{E_2(\si_b)\cdots E_2(\si_{m+1})}\cdot \Pem\sk{E_2(\si_{m})\cdots E_2(\si_1)}\big).
\end{gather*}

To evaluate the projections in  formulas \r{po1} and \r{po2}, we  commute
the negative projections $\Pfm\sk{F_2(s_1)\cdots F_2(s_k)}$ to the left through the product of the
total currents $F_1(t_1)\cdots$ $ F_1(t_a)$ in case of \r{po1} and  commute
the negative projections $\Pem\sk{E_2(s_{m})\cdots E_2(s_1)}$ to the right through the product
of the total currents $E_1(\tau_a)\cdots E_1(\tau_1)$ in \r{po2}.
To perform this commutation we use
\begin{gather}
\Pfm\sk{F_2(s_1)\cdots F_2(s_k)}=
(-1)^k\ F^-_2(s_1;s_2,\ldots,s_k)\cdots F^-_2(s_{k-1};s_k)F^-_2(s_k),\nonumber\\
\Pem\sk{E_2(\si_m)\cdots E_2(\si_1)}=
(-1)^m\ E^-_2(\si_m)E^-_2(\si_{m-1};\si_m)\cdots E^-_2(\si_1;\si_2,\ldots,\si_m),\label{neg}
\end{gather}
and
\begin{gather*}
F_{3,1}(t)F^-_2(s_1;s_2,\ldots,s_k)=
\frac{q^{-1}s_1-qt}{s_1-t} F^-_2(s_1;s_2,\ldots,s_k,t)F_{3,1}(t),\\
E^-_2(\si_1;\si_2,\ldots,\si_m)E_{1,3}(\tau)=
\frac{q^{-1}\si_1-q\tau}{\si_1-\tau}E_{1,3}(\tau)
E^-_2(\si_1;\si_2,\ldots,\si_m,\tau).
\end{gather*}
The expressions
\begin{gather*}
F^-_2(s_1;s_2,\ldots,s_k)=
F^-_2(s_1)-\sum_{\ell=2}^k\frac{s_1}{s_\ell}\ \phi_{s_\ell}(s_1;s_2,\ldots,s_k)F^-_2(s_\ell)
,\\
E^-_2(\si_1;\si_2,\ldots,\si_m)=
E^-_2(\si_1)-\sum_{\ell=2}^m\phi_{\si_\ell}(\si_1;\si_2,\ldots,\si_m)E^-_2(\si_\ell)
\end{gather*}
are linear combinations of the half-currents, while
\begin{gather*}
\phi_{s_\ell}(s_1;s_2,\ldots,s_k)=\prod_{j=2,\, j\neq\ell}^k\frac{s_1-s_j}{s_\ell-s_j}
\prod_{j=2}^k\frac{q^{-1}s_\ell-qs_j}{q^{-1}s_1-qs_j}
\end{gather*}
are rational functions satisfying the normalization conditions
$\phi_{s_j}(s_i;s_2,\ldots,s_k)=\delta_{ij}$, $i,j=2,\ldots,k$.
One also needs the commutation relations between
negative half-currents and the total currents
\begin{gather*}
F_1(t)F^-_2(s)=
\frac{qs-q^{-1}t}{s-t} \sk{F_2^-(s)-\frac{(q-q^{-1})s}{qs-q^{-1}t} F^-_2(t) } F_1(t)
+\frac{s}{t-s}  F_{3,1}(t),\\
E^-_2(\si)E_1(\tau)=
\frac{q\si-q^{-1}\tau}{\si-\tau} E_1(\tau)\sk{E_2^-(\si)-\frac{(q-q^{-1})\tau}{q\si-q^{-1}\tau}
E^-_2(\tau) }
+\frac{\tau}{\si-\tau}\ E_{1,3}(\tau),
\end{gather*}
and the identity
\begin{gather*}
\prod_{i<j}\frac{q^{-1}t_i-qt_j}{t_i-t_j} \, \qSym_{\bar t}\sk{Y(t_n,\ldots,t_1;^{\omega}\!\bar s)}
=\prod_{i<j}\frac{q^{-1}s_i-qs_j}{s_i-s_j}\,  \qSym_{\bar s}\big(Y(^{\omega'}\!\bar t;s_n,\ldots,s_1)\big)
\end{gather*}
valid for arbitrary permutations $\omega$ and $\omega'$ of the sets $\bar s$ and $\bar t$, respectively.

\subsection{Integral presentation of the projections (\ref{po1}) and (\ref{po2})}

The projections \r{po1} and \r{po2} are given as a product of projection of currents.
As already mentioned, this form is not convenient to obtain scalar products.
We give a new
representation in term of a multiple integral over the product of simple root currents:

\begin{proposition}
\begin{gather}
\Pep\sk{E_2(\sigma_b)\cdots E_2(\si_1)E_1(\tau_a)\cdots E_1(\tau_1)}
\nonumber\\
\quad{}=\oint \frac{d\nu_{1}}{\nu_{1}}\cdots \oint \frac{d\nu_{b}}{\nu_{b}}
\oint \frac{d\mu_1}{\mu_1}\cdots \oint \frac{d\mu_a}{\mu_a}\
\KerE(\bar\tau,\bar\si;\bar\mu,\bar\nu)  E_1(\mu_{1})\cdots E_1(\mu_{a})
E_2(\nu_b)\cdots  E_2(\nu_1),
\nonumber\\
\Pfp\sk{F_1(t_1)\cdots F_1(t_a)F_2(s_1)\cdots F_2(s_b)}
\label{NewS}\\
\quad{}=\oint \frac{dy_{1}}{y_{1}}\cdots \oint \frac{dy_{b}}{y_{b}}
\oint \frac{dx_1}{x_1}\cdots \oint \frac{dx_a}{x_a} \KerF(\bar t,\bar s;\bar x,\bar y)
F_2(y_{1})\cdots F_2(y_{b})
F_1(x_{a})\cdots F_1(x_1),
\nonumber
\end{gather}
where the kernels $\KerE(\bar\tau,\bar\si;\bar\mu,\bar\nu)$ and $\KerF(\bar t,\bar s;\bar x,\bar y)$
are given by the series
\begin{gather}
\KerE(\bar\tau,\bar\si;\bar\mu,\bar\nu)= \overline{{\rm Sym}}_{\bar\tau,\bar\si}
\left(\sum_{m=0}^{\min\{a,b\}}
\frac{(q^{-1}-q)^m}{m!(a-m)!(b-m)!}
\prod_{m< i<j\leq b}\frac{\si_i-\si_j}{q^{-1}\si_i-q\si_j}\right.\nonumber\\
 \qquad{} \times\prod_{1\leq i<j\leq a-m\atop a-m<i<j\leq a}\frac{\tau_i-\tau_j}{q^{-1}\tau_i-q\tau_j}
Y(\tau_{a},\ldots,\tau_{a-m+1};\si_m,\ldots,\si_1)
Z(\tau_a,\ldots,\tau_1;\mu_a,\ldots,\mu_1)\nonumber\\
  \left. \qquad{}
\times Z(\mu_a,\ldots,\mu_{a-m+1},\si_{m+1},\ldots,\si_b;\nu_1,\ldots,\nu_b)
\prod_{j=1}^{a-m}\prod_{i=m+1}^b \frac{q^{-1}-q\nu_i/\mu_j}{1-\nu_i/\mu_j}\right)\label{EKer}
\end{gather}
and
\begin{gather}
\KerF(\bar t,\bar s;\bar x,\bar y)= \overline{{\rm Sym}}_{\bar t,\bar s}
\left(\sum_{k=0}^{\min\{a,b\}}
\frac{(q^{-1}-q)^k}{k!(a-k)!(b-k)!}
\prod_{k< i<j\leq b}\frac{s_i-s_j}{q^{-1}s_i-qs_j}\right.\nonumber\\
 \qquad{}\times\prod_{1\leq i<j\leq a-k\atop a-k<i<j\leq a}\frac{t_i-t_j}{q^{-1}t_i-qt_j}
Z(t_{a},\ldots,t_{a-k+1};s_k,\ldots,s_1)
Y(t_a,\ldots,t_1; x_a,\ldots,x_1)\nonumber\\
  \left.\qquad{}
\times Y(x_a,\ldots,x_{a-k+1},s_{k+1},\ldots,s_b;y_1,\ldots,y_b)
\prod_{j=1}^{a-k}\prod_{i=k+1}^b
\frac{q^{-1}-qy_i/x_j}{1-y_i/x_j}\right).\label{FKer}
\end{gather}
\end{proposition}

The proof of these formulas is given in the next subsection.
Let us explain the meaning of the integral formulas for the projections \r{NewS}. There is
a preferable order of integration
in these formulas. First, we have to calculate the integrals over variables $\nu_i$ and $y_i$,
$i=1,\ldots,b$, respectively,
and then calculate the integrals over $\mu_j$ and $x_j$, $j=1,\ldots,a$.

\begin{example}
Let us
illustrate how it works in the simplest
example $a=b=1$ and for projection $\Pfp\sk{F_1(t)F_2(s)}$. We  have
\begin{gather*}
\Pfp\sk{F_1(t)F_2(s)}=\oint \frac{dy}{y}\oint \frac{dx}{x}  \KerF(t,s;x,y)
F_2(y)F_1(x) ,
\end{gather*}
where
\begin{gather*}
 \KerF(t,s;x,y)=Y(t;x)Y(s;y)\frac{q^{-1}-qy/x}{1-y/x}+(q^{-1}-q)Z(t,s)Y(t;x)Y(x;y) .
\end{gather*}
Integration over $y$ with the f\/irst term of the kernel yields to
\begin{gather*}
\oint\frac{dy}{y}  \frac{1}{1-y/s}\frac{q^{-1}-qy/x}{1-y/x}  F_2(y)F_1(x)
=\frac{q^{-1}x-qs}{x-s}F^+_2(s;x)F_1(x)=
F_1(x)F^+_2(s) ,
\end{gather*}
due to the commutation relations \r{us-com}. Integration over $y$ with the second term of the kernel
produces
\begin{gather*}
(q^{-1}-q)F_2^+(x)F_1(x)=\Sfb\sk{F_1(x)} ,
\end{gather*}
according to the formulas~\r{cc-v-so2}. Finally,  integration over $x$ in both terms produces the
result for the projection in this simplest case. The general case can be treated analogously.
Of course, one can  f\/irst integrate over~$x$ and then over~$y$.
However in this case, the calculation of the integrals
for the projection becomes more involved and requires more complicated commutation relations between
half-currents.
\end{example}

\subsection{Proof of the integral presentation of the projections (\ref{po1}) and (\ref{po2})}

Integral representation for the projections of the same type of currents
$\Pfp\sk{F_i(s_{1})\cdots F_i(s_{b})}$ and $\Pep\sk{E_i(\si_{b})\cdots E_i(\si_{1})}$ ($i=1,2$)
were obtained in~\cite{KP}. They can be obtained from the calculation of these
projections
\begin{gather}
\Pfp\sk{F_i(s_{1})\cdots F_i(s_{b})}=
F^+_i(s_1)F^+_i(s_2;s_1)\cdots F^+_i(s_b;s_{b-1},\ldots,s_{1})
,\nonumber\\
\Pep\sk{E_i(\si_{b})\cdots E_i(\si_{1})} =
E^+_i(\si_b;\si_{b-1},\ldots,\si_1)\cdots E^+_i(\si_2;\si_1)E^+_i(\si_1)
,\label{por3}
\end{gather}
where $F^+_i(s_k;s_{k-1},\ldots,s_1)$ and $E^+_i(\si_k;\si_{k-1},\ldots,\si_1)$
are linear combinations of the half-currents
\begin{gather*}
F^+_i(s_k;s_{k-1},\ldots,s_1)=
F^+_i(s_k)-\sum_{\ell=1}^{k-1}\frac{s_k}{s_\ell} \varphi_{s_\ell}(s_k;s_{k-1},\ldots,s_1)F^+_i(s_\ell)
,\\
E^+_i(\si_k;\si_{k-1},\ldots,\si_1)=
E^+_i(\si_k)-\sum_{\ell=1}^{k-1}\varphi_{\si_\ell}(\si_k;\si_{k-1},\ldots,\si_1)F^+_i(\si_\ell),
\end{gather*}
with coef\/f\/icients being rational functions
\begin{gather*}
\varphi_{s_\ell}(s_k;s_{k-1},\ldots,s_1)=\prod_{j=1,\, j\neq\ell}^{k-1}\frac{s_k-s_j}{s_\ell-s_j}
\prod_{j=1}^{k-1}\frac{qs_\ell-q^{-1}s_j}{qs_k-q^{-1}s_j}.
\end{gather*}
There is a very simple analytical proof of the formulas \r{por3} given in~\cite{KP}.

\begin{example}
Let us illustrate this method
on one example: the f\/irst relation in \r{por3} with $b=2$. Indeed,
from the commutation relation of the total currents $F_i(s_1)$ and
$F_i(s_2)$, and due to the integral presentation of negative half-currents
\begin{gather}\label{neg-hc}
F^-_i(s_2)=-\int\frac{dy}{y} \frac{s_2/y}{1-s_2/y}  F_i(y),
\end{gather}
we know that
\begin{gather}\label{hint}
\Pfp\sk{F_i(s_{1})F_i(s_{2})}= F^+_i(s_{1})F^+_i(s_{2})+\frac{s_2}{s_1}
\frac{X(s_1)}{q^{-1}s_1-qs_2} ,
\end{gather}
where $X(s_1)$ is an unknown algebraic element which depends only on the spectral parameter $s_1$.
This element
can be uniquely def\/ined from the relation \r{hint} setting $s_1=s_2$ and using the fact that
$F^2_i(s)=0$. The general case can be treated analogously (see details in \cite{KP}). Formulas~\r{neg}
can be proved  in the same way.
\end{example}

Using now the integral form of the half-currents
\begin{gather*}
F_i^+(s)=\int\frac{dy}{y}\frac{1}{1-y/s}F_i(y),\qquad
E_i^+(s)=\int\frac{dy}{y}\frac{y/s}{1-y/s}E_i(y),
\end{gather*}
one can easily obtain integral formulas for~\r{por3}:
\begin{gather}
 \Pfp\sk{F_i(s_{1})\cdots F_i(s_{b})}=\prod_{1\leq i<j\leq b}\frac{s_i-s_j}{q^{-1}s_i-qs_j}
\nonumber \\
 \qquad{}\times\int \frac{dy_1}{y_1}\cdots \frac{dy_b}{y_b}\
F_i(y_{1})\cdots F_i(y_{b}) Y(s_1,\ldots,s_b;y_1,\ldots,y_b)
 ,\nonumber\\
 \Pep\sk{E_i(\si_{b})\cdots E_i(\si_{1})}=\prod_{1\leq i<j\leq b}\frac{\si_i-\si_j}{q^{-1}\si_i-q\si_j}
\nonumber\\
\qquad{}\times\int \frac{d\nu_1}{\nu_1}\cdots \frac{d\nu_b}{\nu_b}\ E_i(\nu_{b})\cdots E_i(\nu_{1})
Z(\si_1,\ldots,\si_b;\nu_1,\ldots,\nu_b) .\label{por4}
\end{gather}
According to the structure of the kernels \r{rat-Y}, the integrands in \r{por4} have only simple poles
with respect
to the integration variables $y_1$ and $\nu_1$ in the points
$s_1$ and $\si_1$ respectively, while with respect to the variables $y_b$ and $\nu_b$ they have simple
poles
in the points $s_j$ and $\si_j$, $j=1,\ldots,b$. Due to $q$-symmetric prefactors
in the integrals  \r{por4},
the integrals themselves are symmetric with respect to the spectral parameters $s_j$ and $\si_j$,
$j=1,\ldots,b$,
respectively.

The integral form for the projections of the strings
\[
\Pfp\sk{F_1(t_{1})\cdots  F_1(t_{a-k})F_{3,1}(t_{a-k+1})\cdots F_{3,1}(t_{a})}
\]
and
\[
\Pep\sk{ E_{1,3}(\tau_{a})\cdots E_{1,3}(\tau_{a-m+1}) E_1(\tau_{a-m})\cdots  E_1(\tau_{1})  }
\]
is a more delicate question. To present them as integrals, we  use arguments of \cite{KP}
and formulas~\r{cc-v-so2}. The point is that the analytical properties of the reverse strings
\[
\Pfp\sk{F_{3,1}(t_{a})\cdots  F_{3,1}(t_{a-k+1})F_1(t_{a-k})\cdots F_1(t_{1})}
\]
 and
\[
\Pep\sk{E_1(\tau_{1}) \cdots  E_1(\tau_{a-m})E_{1,3}(\tau_{a-m+1})\cdots  E_{1,3}(\tau_{a})  }
\]
are the same as the analytical properties of the product of the simple root currents
$F_{1}(t_{a}){\cdots} F_1(t_{1})$ and $E_1(\tau_{1}) {\cdots}   E_{1}(\tau_{a})$. Therefore the calculation of
projection of the reverse string can be done along the same steps as for the product of simple root
currents. In order to relate the projection of the string and projection of the reverse string, we
need the commutation relations
\begin{gather*}
F_1(t_1)F_{3,1}(t_2)=\frac{qt_1-q^{-1}t_2}{t_1-t_2}\ F_{3,1}(t_2)F_1(t_1),\nonumber\\
E_{1,3}(\tau_2)E_1(\tau_1)=\frac{q\tau_1-q^{-1}\tau_2}{\tau_1-\tau_2}\ E_1(\tau_1)E_{1,3}(\tau_2)
\end{gather*}
and the fact (proved in \cite{KP}) that under projections we can freely
exchange currents without
taking into account the $\delta$-function terms. As result, we get
\begin{gather}
\Pfp\sk{F_1(t_{1})\cdots  F_1(t_{a-k})F_{3,1}(t_{a-k+1})\cdots F_{3,1}(t_{a})}
\nonumber\\
 \qquad{}=
\prod_{1\leq i\leq a-k\atop a-k<j\leq a}\frac{qt_i-q^{-1}t_j}{t_i-t_j}
\prod_{1\leq i<j\leq a-k\atop a-k<i<j\leq a}\frac{qt_i-q^{-1}t_j}{q^{-1}t_i-qt_j}
\nonumber\\
 \qquad\quad{} \times \prod^{\longleftarrow}_{a\geq\ell>a-k}\Pfp\sk{F_{3,1}(t_\ell;t_{\ell+1},\ldots,t_a)}
\prod^{\longleftarrow}_{a-k\geq\ell\geq 1}F^+_{1}(t_\ell;t_{\ell+1},\ldots,t_a)
\nonumber\\
 \qquad {} = \prod_{1\leq i<j\leq a-k\atop a-k<i<j\leq a}\frac{t_i-t_j}{q^{-1}t_i-qt_j}
\int \frac{dx_1}{x_1}\cdots \frac{dx_a}{x_a}\ Y(t_a,\ldots,t_1;x_a,\ldots,x_1)
\nonumber\\
 \qquad\quad{}\times \Sfb(F_1(x_a))\cdots  \Sfb(F_1(x_{a-k+1})) F_1(x_{a-k})\cdots F_1(x_1).\label{Pstf}
\end{gather}
Analogously
\begin{gather}
\Pep\sk{ E_{1,3}(\tau_{a})\cdots E_{1,3}(\tau_{a-m+1}) E_1(\tau_{a-m})\cdots  E_1(\tau_{1})  }\nonumber\\
 \qquad{}=\prod_{1\leq i\leq a-k\atop a-k<j\leq a}\frac{q\tau_i-q^{-1}\tau_j}{\tau_i-\tau_j}
\prod_{1\leq i<j\leq a-k\atop a-k<i<j\leq a}\frac{q\tau_i-q^{-1}\tau_j}{q^{-1}\tau_i-q\tau_j}
\nonumber\\
 \qquad\quad{} \times \prod^{\longrightarrow}_{1\leq\ell\leq a-m} E^+_{1}(t_\ell;t_{\ell+1},\ldots,t_a)
\prod^{\longrightarrow}_{a-k<\ell\leq a}  \Pep\sk{E_{1,3}(t_\ell;t_{\ell+1},\ldots,t_a)} \nonumber\\
 \qquad {} = \prod_{1\leq i<j\leq a-m\atop a-m<i<j\leq a}\frac{\tau_i-\tau_j}{q^{-1}\tau_i-q\tau_j}
\int \frac{d\mu_1}{\mu_1}\cdots \frac{d\mu_a}{\mu_a}  Z(\tau_a,\ldots,\tau_1;\mu_a,\ldots,\mu_1)
\nonumber\\
 \qquad\quad{}\times E_1(\mu_{1})\cdots E_1(\mu_{a-m}) \Seb(E_1(\mu_{a-m+1}))\cdots  \Seb(E_1(\mu_{a}))
 .\label{Pste}
\end{gather}

Here we used the notations
\[
\prod^{\longleftarrow}_{a\geq\ell\geq 1} A_\ell = A_a A_{a-1}\cdots A_2A_1,\qquad
\prod^{\longrightarrow}_{1\leq\ell\leq a} B_\ell= B_1B_2\cdots B_{a-1} B_a
\]
for products of  non-commutative terms and the identities
\begin{gather*}
\Pfp\sk{F_{3,1}(t)} =\Pfp\sk{\Sfb(F_1(t))}=\Sfb\sk{\Pfp(F_1(t))}=\Sfb\sk{F^+_1(t)}
 ,\\
\Pep\sk{E_{1,3}(\tau)} =\Pep\sk{\Seb(E_1(\tau))}=\Seb\sk{\Pep(E_1(\tau))}=\Seb\sk{E^+_1(\tau)}
 ,
\end{gather*}
on commutativity of the screening operators and the projections proved in \cite{KP}.

The last step before getting integral formulas for universal Bethe vectors is to present products
of  screening operators acting on total currents,
\begin{gather*}
\Sfb(F_1(x_k))\cdots  \Sfb(F_1(x_{1})),\qquad
\Seb(E_1(\mu_{1}))\cdots  \Seb(E_1(\mu_{m})),
\end{gather*}
 as an integral using formulas \r{cc-v-so2}.
The presentation follows from the following chain of equa\-li\-ties
\begin{gather}
\Sfb(F_1(x_k))\cdots  \Sfb(F_1(x_{1}))=
(q^{-1}-q)^k F^+_2(x_k)F_1(x_k)\cdots F^+_2(x_2)F_1(x_2)F^+_2(x_1)F_1(x_1)\nonumber\\
 =(q^{-1}\! -q)^k \!\!\! \prod_{1\leq i<j\leq k} \!\!\! \frac{qx_i\!-\!q^{-1}x_j}{x_i-x_j}
F_2^+(x_k)F_2^+(x_{k-1};x_k)\cdots F_2^+(x_1;x_2,\dots ,x_k)
F_1(x_k)\cdots F_1(x_1)\nonumber\\
 =(q^{-1}-q)^k \int \frac{dz_1}{z_1}\cdots  \frac{dz_z}{z_k} Y(x_k,\dots,x_1;z_k,\dots,z_1)
 F_2(z_k)\cdots F_2(z_1)F_1(x_k)\cdots F_1(x_1)\!\!\!\label{deqf}
\end{gather}
and
\begin{gather}
\Seb(E_1(\mu_{1}))\cdots  \Seb(E_1(\mu_{m}))=
(q^{-1}-q)^k  E_1(\mu_1)E^+_2(\mu_1)\cdots E_1(\mu_m)E^+_2(\mu_m)\nonumber\\
 =(q^{-1}-q)^m \prod_{1\leq i<j\leq m} \frac{q\mu_i-q^{-1}\mu_j}{\mu_i-\mu_j}
 E_1(\mu_1)\cdots  E_1(\mu_m) E^+_2(\mu_1;\mu_2,\dots,\mu_m)\cdots E^+_2(\mu_m)\!\!\!\label{deqe}\\
 = (q^{-1}-q)^m \int \frac{d\rho_1}{\rho_1}\cdots
\frac{d\rho_z}{\rho_m} Z(\mu_m,\dots,\mu_1;\rho_m,\dots,\rho_1)
E_1(\mu_1)\cdots  E_1(\mu_m)E_2(\rho_1)\cdots  E_2(\rho_m) ,\nonumber
\end{gather}
where we have used the commutation relation
\begin{gather}
F_1(x_j)F^+_2(x_i;x_{i-1},\ldots,x_{j-1})
=\frac{qx_i-q^{-1}x_j}{x_i-x_j} F^+_2(x_i;x_{i-1},\ldots,x_{j})
F_1(x_j) ,\nonumber\\
E^+_2(\mu_i;\mu_{i+1},\ldots,\mu_{j-1})E_1(\mu_j) =\frac{q\mu_i-q^{-1}\mu_j}{\mu_i-\mu_j}
E_1(\mu_j)E^+_2(\mu_i;\mu_{i+1},\ldots,\mu_{j}) .\label{us-com}
\end{gather}
Note that these commutation formulas are crucial for the integral formulas
 given below in \r{Fie1} and \r{Fif1}.
 One can see that the right hand sides of these formulas are not ordered,
while the left hand sides are.
\begin{example}
Let us check the f\/irst equality in \r{us-com}, in the simplest case.
To calculate this exchange relation, we start
from the def\/inition of the composed currents $F_{3,1}(x)$
as given in~\r{delta-terms}
and apply to this relation the integral transformation
\begin{gather*}
\int \frac{dy}{y} \frac{1}{1-y/x_1}.
\end{gather*}
To calculate this integral, we decompose the kernel of the integrand
as
\begin{gather*}
\frac{q-q^{-1}x_2/y}{1-x_2/y}\cdot \frac{1}{1-y/x_1}=
\frac{q-q^{-1}x_2/x_1}{1-x_2/x_1} \cdot \frac{1}{1-y/x_1}+
     \frac{(q-q^{-1})}{1-x_2/x_1}\cdot \frac{x_2/y}{1-x_2/y} .
\end{gather*}
This leads to
\begin{gather*}
F_{1}(x_2)F^+_{2}(x_1)  =
  \frac{q-q^{-1}x_2/x_1}{1-x_2/x_1}  F^+_{2}(x_1)F_{1}(x_2)
\nonumber\\
\phantom{F_{1}(x_2)F^+_{2}(x_1)  =}{} - \frac{(q-q^{-1})}{1-x_2/x_1}  F^-_{2}(x_2)F_{1}(x_2)
+\frac{1}{1-x_2/x_1}F_{3,1}(x_2)\nonumber\\
\phantom{F_{1}(x_2)F^+_{2}(x_1)   }{}  =  \frac{q-q^{-1}x_2/x_1}{1-x_2/x_1}
\sk{F^+_{2}(x_1)-\frac{(q-q^{-1})x_1}{q-q^{-1}x_2/x_1}  F^+_{2}(x_2)}F_{1}(x_2)\nonumber\\
\phantom{F_{1}(x_2)F^+_{2}(x_1)   }{}  = \frac{q-q^{-1}x_2/x_1}{1-x_2/x_1}  F^+_{2}(x_1;x_2)F_{1}(x_2),
\end{gather*}
where we have used the def\/inition of the negative half-current \r{neg-hc}, the
expression of the
total composed current \r{com-as-prod} and the Ding--Frenkel relation
$F_2(x_2)=F^+_2(x_2)-F_2^-(x_2)$.
\end{example}

After substituting formulas \r{deqf} and \r{deqe} into integral formulas for the projections of the
string \r{Pstf} and  \r{Pste}, we obtain, from the resolution of the hierarchical relations for the
universal Bethe vectors \r{po1} and \r{po2}, the following intermediate results
\begin{gather}
 \Pep\sk{E_2(\sigma_b)\cdots E_2(\si_1)E_1(\tau_a)\cdots E_1(\tau_1)}= \sum_{m=0}^{\min\{a,b\}}
\frac{(q^{-1}-q)^m}{m!(a-m)!(b-m)!}\nonumber\\
\qquad{}\times \overline{{\rm Sym}}_{\bar\tau,\bar\si}
\left(
\prod_{1\leq i<j\leq a-m\atop a-m<i<j\leq a}\frac{\tau_i-\tau_j}{q^{-1}\tau_i-q\tau_j}
Y(\tau_{a},\ldots,\tau_{a-m+1};\si_m,\ldots,\si_1) \right.\nonumber\\
\qquad{}\times
\oint \frac{d\nu_{1}}{\nu_{1}}\cdots \oint \frac{d\nu_{m}}{\nu_{m}}
\oint \frac{d\mu_1}{\mu_1}\cdots \oint \frac{d\mu_a}{\mu_a}\nonumber\\
\qquad{}\times
 Z(\tau_a,\ldots,\tau_1;\mu_a,\ldots,\mu_1)
Z(\mu_a,\ldots,\mu_{a-m+1};\nu_1,\ldots,\nu_m)\nonumber\\
\left.\vphantom{\prod_{a\atop a}}\qquad{}\times
\Pep\sk{E_2(\si_{b})\cdots E_2(\si_{m+1})}\ E_1(\mu_{1})\cdots E_1(\mu_{a})
E_2(\nu_m)\cdots  E_2(\nu_1)\right)\label{Fie1}
\end{gather}
and
\begin{gather}
 \Pfp\sk{F_1(t_1)\cdots F_1(t_a)F_2(s_1)\cdots F_2(s_b)}= \sum_{k=0}^{\min\{a,b\}}
\frac{(q^{-1}-q)^{k}}{k!(a-k)!(b-k)!}\nonumber\\
\qquad {} \times\overline{{\rm Sym}}_{\bar t,\bar s}
\left(\prod_{1\leq i<j\leq a-k\atop a-k<i<j\leq a}\frac{t_i-t_j}{q^{-1}t_i-qt_j}
  Z(t_{a},\ldots,t_{a-k+1};s_k,\ldots,s_1)\right.\nonumber\\
 \qquad{} \times
\oint \frac{dy_{1}}{y_{1}}\cdots \oint \frac{dy_{k}}{y_{k}}
\oint \frac{dx_1}{x_1}\cdots \oint \frac{dx_a}{x_a}\nonumber\\
\qquad{} \times  Y(t_a,\ldots,t_1;x_a,\ldots,x_1)
Y(x_a,\ldots,x_{a-k+1};y_1,\ldots,y_k)\nonumber\\
 \left.\qquad\vphantom{\prod_{a\atop a}}{}\times
F_2(y_{1})\cdots F_2(y_{k})
F_1(x_{a})\cdots F_1(x_1)
\Pfp\sk{F_2(s_{k+1})\cdots F_2(s_{b})}\right) .\label{Fif1}
\end{gather}
The last step is to move to the left, in \r{Fie1}, the product of the total currents
$E_1(\mu_{1})\cdots$ $ E_1(\mu_{a})$
through the projection $\Pep\sk{E_2(\si_{b})\cdots E_2(\si_{m+1})}$ using the factorization formulas
\r{por3} and the commutation relations \r{us-com}.
Analogously, in \r{Fif1}, one has to move to the right
 the product of the total currents $F_1(x_{a})\cdots F_1(x_1)$
through the projection $\Pfp\sk{F_2(s_{k+1})\cdots F_2(s_{b})}$, using again
the factorization formulas
\r{por3} and the commutation relations \r{us-com}.  As result, we  obtain the
integral formulas \r{NewS} for the projections of the product of currents for the algebra $\Uqtri$.

\section{Scalar products of universal Bethe vectors}

\subsection{Commutation of products of total currents}

Formulas \r{NewS} show that in order to calculate the scalar product of the universal Bethe
vectors, one has to commute the products of the total currents
\[
\CalE(\bar\mu,\bar\nu)= E_1(\mu_{1})\cdots E_1(\mu_{a})\ E_2(\nu_b)\cdots  E_2(\nu_1)
\]
and
\[
\CalF(\bar x,\bar y)=F_2(y_{1})\cdots F_2(y_{b})\ F_1(x_{a})\cdots F_1(x_1).
\]

According to the decomposition of the quantum af\/f\/ine algebra $\Uqtri$
used in this paper, the modes of the total currents
$F_i[n]$, $E_i[n+1]$, $k_j^+[n]$, $n\geq0$ and a $q$-commutator
$E_{1,3}[1]=E_2[0]E_1[1]-q^{-1}E_1[1]E_2[0]$,
 belong to the Borel subalgebra
$\Uqbp\in\Uqtri$. We def\/ine the following ideals in this Borel subalgebra.

\begin{definition}\label{def:J}
We note $J$, the left ideal of $\Uqbp$ generated by all elements of the form
$\Uqbp\cdot E_i[n]$, $n>0$ and $\Uqbp\cdot E_{1,3}[1]$. Equalities in $\Uqbp$ modulo element from the ideal
$J$ are denoted by the symbol `$\simJ$'.
\end{definition}

\begin{definition}\label{def:I}
Let $I$ be the right ideal of $\Uqbp$ generated by all elements of the form
$F_i[n]\cdot \Uqbp$ such that  $n\geq0$. We denote equalities modulo
elements from the ideal $I$ by the symbol~`$\simI$'.
\end{definition}

We also def\/ine the following ideal in $\Uqtri$:

\begin{definition}\label{def:K}
We denote by $K$ the two-sided $\Uqtri$ ideal
  generated by the elements which have at least one
arbitrary mode $k_j^-[n]$, $n\leq0$, of the negative Cartan
current $k_j^-(t)$. Equalities in $\Uqtri$ modulo element of the ideal $K$ are denoted by
the symbol `$\simK$'.
\end{definition}

Equalities in $\Uqtri$ modulo the right ideal $I$, the left ideal $J$
and the two-sided ideal $K$ will be denoted by the
symbol `$\simm$'.

A right weight singular vector def\/ined by the relations \r{hwv} is
annihilated by the right action of  any positive  mode  $E_i[n]$,
$n> 0$, the element $E_{1,3}[1]$ and is a right-eigenvector for  $k_j^+(t)$,
\begin{gather*}
E_i^+(\tau)\cdot v=0 ,\qquad  \Pep\sk{E_{1,3}(\tau)}\cdot v=0,\qquad k_j^+(\tau)\cdot v
=\Lambda_j(\tau) v ,
\end{gather*}
where $\Lambda_j(\tau)$ are some meromorphic functions, decomposed as a
power series in $\tau^{-1}$.
A left weight singular vector $v'$ def\/ined by the relation~\r{hwvl}   is
annihilated by the left action of any nonnegative  modes  $F_i[n]$,
$n\geq 0$ and is a left-eigenvector for  $k^+_j(t)$,
\begin{gather*}
v'\cdot F_i^+(t)=0  ,\qquad v'\cdot k_j^+(t)=\Lambda'_j(t) v'  ,
\end{gather*}
where $\Lambda'_j(t)$ are also meromorphic functions.
These facts follow from the relation between projections of the currents and
the Gauss coordinates of the $\LL$-operator \r{GF2}--\r{GE2}.

We observe that the vectors
\begin{gather}\label{unbv1}
\Pfp\sk{F_1(t_1)\cdots F_1(t_a)F_2(s_1)\cdots F_2(s_b)}\cdot v
\end{gather}
and
\begin{gather}\label{unbv2}
v'\cdot \Pep\sk{E_2(\sigma_b)\cdots E_2(\si_1)E_1(\tau_a)\cdots E_1(\tau_1)}
\end{gather}
belong to the modules over the
quantum af\/f\/ine algebra $\Uqtri$ from
the categories of the highest weight and lowest weight representations respectively.
This is in accordance with the def\/inition of the completions $\overline U_E$ and
$\overline U_F$ and the corresponding
projections given above.

We assume the existence of a nondegenerate pairing $\<v',v\>$ and by the scalar product
of the left and right universal Bethe vectors, we will understand the coef\/f\/icient
$\Scp(\bar\tau,\bar\si;\bar t,\bar s)$ in front of the pairing $\<v',v\>$
in the right hand side of equality
\begin{gather}
 \<v'\cdot \Pep\sk{E_2(\sigma_b)\cdots E_2(\si_1)E_1(\tau_a)\cdots E_1(\tau_1)},
\Pfp\sk{F_1(t_1)\cdots F_1(t_a)F_2(s_1)\cdots F_2(s_b)}\cdot v\>
\nonumber\\
 \qquad{}=\Scp(\tau_1,\ldots,\tau_a,\si_1,\ldots,\si_b;
t_1,\ldots,t_a,s_1,\ldots,s_b)\<v',v\>.\label{sc-pr}
\end{gather}
It is clear that the scalar product \r{scal-prod} dif\/fers from \r{sc-pr} by the product
\[
\prod_{k=1}^a \Lambda_1(t_k)\Lambda'_1(\tau_k) \prod_{m=1}^b \Lambda_2(s_m)\Lambda'_2(\si_m)  .
\]

The problem of calculation of the scalar product of the universal Bethe vectors
\r{sc-pr} is equiva\-lent to the commutation of the projections entering
the def\/initions of the vectors~\r{unbv1} and~\r{unbv2}
modulo the left ideal $J$ and the right ideal~$I$.
To calculate this commutation, we  use the integral presentation of the projections
\r{NewS},  commute  the total currents and then calculate the integrals.
Since both projections belong to the positive Borel subalgebra~$\Uqbp$, we can
neglect the terms which contain the negative Cartan currents~$k_i^-(t)$ and perform
the commutation of the total currents modulo
the two-sided ideal~$K$. Actually, in commuting the total currents, we
will be interested only in terms
which are products of combinations of the $\Uqtri$ positive Cartan currents~\r{Car-com}.
All other terms will be annihilated by the weight singular vectors.

Let us recall that elements $\CalE(\bar\mu,\bar\nu)$ and
$\CalF(\bar x,\bar y)$ are elements of the completed algebras $\overline U_E$ and $\overline U_F$,
which are dual subalgebras in $\Uqtri$ considered as a quantum double. There is a~nondegenerate
Hopf pairing between these subalgebras, given by the formulas \r{Hopf-pair}.
For any elements $a\in\mathcal{A}$ and $b\in\mathcal{B}$ from two dual Hopf subalgebras
$\mathcal{A}$ and $\mathcal{B}$ of the
quantum double algebra
$\mathcal{D}(\mathcal{A})=\mathcal{A}\oplus\mathcal{B}$, there is a relation~\cite{EKhP}
\begin{gather}\label{com-qd}
\<a^{(2)},b^{(2)}\>\ b^{(1)}\cdot a^{(1)}= a^{(2)}\cdot b^{(2)}  \< a^{(1)},b^{(1)}\>
 ,
\end{gather}
where $\Delta_{\mathcal{A}}(a)=a^{(1)}\otimes a^{(2)}$ and
$\Delta_{\mathcal{B}}(b)=b^{(1)}\otimes b^{(2)}$.

Let us apply formula \r{com-qd} for $a=\CalE(\bar\mu,\bar\nu)=\CalE$
and $b=\CalF(\bar x,\bar y)=\CalF$. Using the current  coproduct \r{gln-copr},
we conclude that
\begin{gather}\label{EF-co}
\Delta^{(D)}\CalE=1\otimes\CalE+\CalE'\otimes \CalE'',\qquad
\Delta^{(D)}\CalF=\mathcal{K}^+\otimes\CalF+\CalF'\otimes \CalF'',
\end{gather}
where the element $\CalE'$  satisf\/ies $\coun(\CalE')=0$ and the element $\CalE''$ contains at least
one negative Cartan current $k^-_i(\tau)$. The element $\mathcal{K}^+$ in \r{EF-co}
takes the form
\begin{gather*}
\mathcal{K}^+=\prod_{i=1}^a \psi_1^+(x_i) \prod_{j=1}^b \psi_2^+(y_j) .
\end{gather*}
The left hand side of the relation \r{com-qd}  have the form
\begin{gather*}
\<\CalE,\CalF\> \cdot \mathcal{K}^+\quad {\rm mod}\; \tilde J
\end{gather*}
and the right hand side of the same relation is
\begin{gather*}
\CalE\cdot \CalF \quad{\rm mod}\; K .
\end{gather*}
The ideal $\tilde J$, similar to the ideal $J$, is the left ideal in $\Uqtri$ generated by
the elements $\Uqtri\cdot E_i[n]$, $i=1,2$ and $n\in\mathbb{Z}$. One can check that after integration
in \r{NewS} the terms of the ideal $\tilde J$ which have non-positive modes of the currents $E_1(\mu_k)$
and $E_2(\nu_m)$ on the right will disappear and can be neglected.
Alternatively,  we can argue that these terms are irrelevant using cyclic ordering of the current or
Cartan--Weyl generators, as it was done in the papers \cite{EKhP,FKPR}.

As result, a general equality
\r{com-qd} for the given elements $a=\CalE(\bar\mu,\bar\nu)$
and $b=\CalF(\bar x,\bar y)$ reads
\begin{gather*}
\CalE(\bar\mu,\bar\nu)\cdot \CalF(\bar x,\bar y) =
\<\CalE(\bar\mu,\bar\nu),\CalF(\bar x,\bar y)\>\prod_{i=1}^a \psi_1^+(x_i)
\prod_{j=1}^b \psi_2^+(y_j)\quad
{\rm mod}\; (K,J)
\end{gather*}
modulo ideals $K$ and $J$. This relation shows that instead of calculating the exchange relations
for the product of the currents $\CalE(\bar\mu,\bar\nu)$
and $\CalF(\bar x,\bar y)$ it is enough to calculate the pairing between them.

\subsection{Pairing and integral formula for scalar products}

To calculate the pairing, we will use the basic properties of pairing between dual Hopf subalgebras
\begin{gather*}
\<a_1a_2,b\>=\<a_1\otimes a_2,\Delta_{\mathcal{B}}(b)\> ,\qquad
\<a,b_1b_2\>=\<\Delta_{\mathcal{A}}(a),b_2\otimes b_1\> ,
\end{gather*}
where $\mathcal{A}=U_E$ and $\mathcal{B}=U_F$.
From these properties, we obtain
\begin{gather*}
\<\CalE(\bar\mu,\bar\nu),\CalF(\bar x,\bar y)\> =
\prod_{i=1}^b\prod_{j=1}^a \frac{q^{-1}x_j-qy_i}{x_j-y_i}
\prod_{i<j}^a \frac{q^{-1}x_i-qx_j}{qx_i-q^{-1}x_j}
\prod_{i<j}^b \frac{q^{-1}y_i-qy_j}{qy_i-q^{-1}y_j}\nonumber\\
\phantom{\<\CalE(\bar\mu,\bar\nu),\CalF(\bar x,\bar y)\> =}{}\times (q-q^{-1})^{a+b}\  \qSym_{\ \bar x}\sk{\prod_{i=1}^a\delta (\mu_i/x_i)}
\qSym_{\bar y}\sk{\prod_{i=1}^b\delta (\nu_i/y_i)} .
\end{gather*}

Using the def\/inition of the scalar product of the universal
Bethe vectors \r{sc-pr} and integral presentations of the projections~\r{NewS},
we conclude
\begin{proposition}
\begin{gather}
 \Scp(\tau_1,\ldots,\tau_a,\si_1,\ldots,\si_b;t_1,\ldots,t_a,s_1,\ldots,s_b) \nonumber\\
\qquad{} =(q-q^{-1})^{a+b}\oint \frac{dx_1}{x_1}\cdots \oint \frac{dx_a}{x_a}
\oint \frac{dy_{1}}{y_{1}}\cdots \oint \frac{dy_{b}}{y_{b}}
\prod_{i=1}^b\prod_{j=1}^a \frac{q^{-1}x_j-qy_i}{x_j-y_i}\nonumber\\
 \qquad\quad{}\times\prod_{i<j}^a \frac{q^{-1}x_i-qx_j}{qx_i-q^{-1}x_j}
\prod_{i<j}^b \frac{q^{-1}y_i-qy_j}{qy_i-q^{-1}y_j}
\KerE(\bar\tau,\bar\si;\bar x,\bar y)
\qSym_{\bar x,\bar y}\big(\KerF(\bar t,\bar s;\bar x,\bar y)\big)\nonumber\\
\qquad\quad {}\times \prod_{i=1}^a \psi_1^+(x_i) \prod_{j=1}^b \psi_2^+(y_j),\label{sc1}
\end{gather}
where the rational series $\KerE(\bar\tau,\bar\sigma;\bar x,\bar y)$ and
$\KerF(\bar t,\bar s;\bar x,\bar y)$ are given in \eqref{EKer} and \eqref{FKer}.
\end{proposition}

\section{Conclusions}

The kernels entering the formulas \r{NewS} can be $q$-symmetrized over integration variables due to the
$q$-symmetric properties of the product of the total currents. In the $\mathfrak{gl}_2$ case, this leads
to the determinant representation of the kernel due to the identity
\begin{gather*}
\prod_{i<j}^n
\frac{q^{-1}t_i-qt_j}{t_i-t_j}  \, \qSym_{\bar t}\sk{Y(\bar t,\bar x)}=
\prod_{i<j}^n
\frac{q^{-1}x_i-qx_j}{x_i-x_j}  \, \qSym_{\bar x}\sk{Y(\bar t,\bar x)}\\
\qquad{} =\frac{\prod_i t_i\prod_{i,j} (q^{-1}t_i-qx_j)  }{\prod_{i<j}(t_i-t_j)(x_j-x_i)}
\det\left|\frac{1}{(t_i-x_j)(q^{-1}t_i-qx_j)}
\right|_{i,j=1,\ldots,n} ,
\end{gather*}
where the determinant on the right hand side is called
an Izergin determinant. It is
equal (up to a scalar factor) to the partition function of the $XXZ$ model with domain wall
boundary conditions~\cite{I-dwbc}.

The challenge is to get determinant formulas for the $q$-symmetrized kernels \r{EKer}  and \r{FKer}
as a sum of determinants and to use further this determinant formula to get
a determinant formula for the scalar products. Work in this direction is in progress.

\subsection*{Acknowledgement}

Authors would like to acknowledge very useful discussions with Sergei Khoroshkin and Nikita Slavnov.
This work was partially done when the second
author (S.P.) visited Laboratoire d'An\-necy-Le-Vieux de Physique Th\'eorique in February, 2009.
This visit was possible due to the f\/inancial support of
the CNRS-Russia exchange program on mathematical physics.
He thanks LAPTH for the hospitality and stimulating scientif\/ic atmosphere.
Work  of S.P.  was supported in part by RFBR
grant 08-01-00667, RFBR-CNRS grant 07-02-92166-CNRS and grant of the Federal Agency for
Science and Innovations of Russian Federation
under contract 14.740.11.0347. Work of S.B. was supported in part by the INFN Iniziativa Specif\/ica FI11.

\pdfbookmark[1]{References}{ref}
\LastPageEnding

\end{document}